# Büyük Veri Teknolojileri ile Bilimsel Makalelerin Sınıflandırılması

## Classification of Scientific Papers With Big Data Technologies

Selen GURBUZ  
*Firat University, Computer Engineering Department*  
*23100, Elazig, Turkey*  
seseng@firat.edu.tr

Galip AYDIN  
*Firat University, Computer Engineering Department*  
*23100, Elazig, Turkey*  
gaydin @firat.edu.tr

*Abstract* - Data sizes that cannot be processed by conventional data storage and analysis systems are named as Big Data. It also refers to new technologies developed to store, process and analyze large amounts of data. Automatic information retrieval about the contents of a large number of documents produced by different sources, identifying research fields and topics, extraction of the document abstracts, or discovering patterns are some of the topics that have been studied in the field of big data. In this study, the Naïve Bayes classification algorithm, which is run on a data set consisting of scientific articles, has been tried to automatically determine the classes to which these documents belong. We have developed an efficient system that can analyze the Turkish scientific documents with the distributed document classification algorithm run on the Cloud Computing infrastructure. The Apache Mahout library is used in the study. The servers required for classifying and clustering distributed documents are

## I. GİRİŞ

Bilişim dünyasındaki büyük gelişmeler, her alanda üretilen verinin miktarının da çok hızlı bir şekilde artmasına sebep olmuştur. İnternetin yaygınlaşması ve Web 2.0 ile beraber sosyal medya ve kullanıcılar arasında paylaşımın artması, çevrimiçi yayın yapan gazete, dergi, blog gibi kaynaklar aracılığıyla yayınlanan bilgi miktarını daha önce görülmedik boyutlara çıkarmıştır. Benzer şekilde bilim dünyasında önceki yıllarla karşılaştırılmayacak sayılarda yeni yayınlar yapılmakta ve dokümanlar üretilmektedir.

Ancak veri miktarındaki artış bu verileri anlamlandırabilecek, ne ile ilgili olduğunu anlayabilecek, ait oldukları sınıfları bulabilecek algoritmaların geliştirilmesi mümkün olduğunda kıymetli olabilmektedir. İşlenmeyen ve bilgiye dönüştürülemeyen veriden faydalanılamaz.

Bilgisayar Bilimlerinin Veri Madenciliği ve Makine Öğrenmesi gibi dalları veri analizi ile ilgili önemli yöntemler ve algoritmalar sunmaktadır. Ancak bu algoritmalar geleneksel olarak çok büyük miktarda veriyi işlemek için tasarlanmamışlardır. Bir algoritmanın işleyebileceği veri miktarı donanımsal kaynakların kapasitesi ile yakından ilgilidir. Çok büyük analizler yapabilmek için işlem ve depolama kapasitesi yüksek bilgisayarlara ihtiyaç duyulmaktadır. Ancak yüksek kapasiteli bilgisayarlar yüksek maliyetleri nedeniyle çok az sayıda bulunabilmekte ve bunlara araştırmacıların önemli bir kısmının erişim imkanı olamamaktadır.

Geleneksel hesaplama yöntemleriyle analiz edilemeyecek boyutlardaki verileri ifade etmek için Büyük Veri (Big Data) kavramı kullanılmaktadır. Son yılların en popüler konularından birisi olan Büyük Veri, Bulut Bilişim, Paralel Hesaplama, MapReduce gibi teknolojileri kullanarak kolaylıkla bulunabilecek, pahalı olmayan bilgisayarlar üzerinde çok büyük verilerin analizine olanak sağlamaktadır.

Bulut Bilişim donanımsal ve yazılımsal kaynakların servis olarak sunularak geleneksel istemci sunucu mimarilerine göre çok daha yüksek ölçeklenebilirliğe ulaşılmasını sağlayan modern bir yaklaşımdır. Bulut Bilişim sayesinde ortalama özellikli sunucular üzerinde çok sayıda sanal sunucu çalıştırılabilmekte ve donanımsal kaynakların daha optimize kullanılması sağlanmaktadır.

Doküman analizinde çok kullanılan metin madenciliği, metinden yüksek kaliteli bilginin çıkartımı işlemi olarak tanımlanabilir. Metin madenciliği işlemleri arasında metin sınıflandırma, metin kümeleme, konu çıkartımı, duygu analizi, doküman özetleme ve ilişki keşfi gibi konular bulunmaktadır.

## II. ARKA PLAN

### A. Dağıtık Sistemler

Dağıtık sistemlerin birçok tanımı bulunmasına rağmen en genel haliyle kullanıcılarına tek bir bilgisayarmış gibi görünen ancak birbirinden bağımsız bilgisayarlardan oluşan sistem olarak tanımlanabilir [1].

Dağıtık sistemlerin bazı temel özellikleri şunlardır: bu sistemler çeşitli bilgisayarların çoğunlukla kullanıcılardan gizli olarak, nasıl haberleştikleri arasındaki ayrımı yapabilmektedir. Başka bir önemli özellik de kullanıcılar ve uygulamalar bir dağıtık sistem ile tutarlı ve değişmeyen bir yolla, nerede ve ne zaman olursa olsun karşılıklı etkileşimde bulunabilirler.

Prensipte, dağıtık sistemlerin geliştirilebilir ve ölçeklenebilir olması istenir. Bu karakteristik özellik, bağımsız bilgisayarlara sahip olmanın ve bu bilgisayarların bir bütün olarak sisteme dahil olmalarının kullanıcılardan gizlenmesinin doğrudan bir sonucudur. Bazı kısımları geçici olarak kullanım dışı olsa bile, dağıtık sistemler normal ve sürekli bir şekilde etkin durumdadır. Kullanıcılar ve uygulamalar sistemdeki değişim ve onarımları görmemeli ve daha fazla kullanıcı ve uygulamaya hizmet edebilmek için yeni kaynakların eklendiği gösterilmemelidir.

Tek sistem görünümünü sunmanın yanında, heterojen bilgisayar ve ağları desteklemek için, dağıtık sistemler genellikle bir yazılım katmanı vasıtasıyla düzenlenir. Bu yazılım mantıksal olarak kullanıcı ve uygulamalardan oluşan yüksek seviyeli bir katman ile işletim sistemleri ve orta katman yazılımı denilen temel iletişim hizmetleri içeren katmanlardan oluşur. İdeal bir dağıtık sistemin özellikleri arasında heterojenlik, ölçeklenebilirlik, güvenli olma ve hata toleransı bulunur.

## B. Hadoop

Hadoop, terabaytlarla ifade edilen ya da daha büyük verileri işlemek amacı ile oluşturulmuş bir kütüphanedir. Hadoop Distributed File System (HDFS) adı verilen dağıtık dosya sistemine sahip olup Java ile geliştirilmiştir. Hadoop'ta uygulama yaparken kullanılan dosya sistemi HDFS (Hadoop File System) olarak adlandırılmıştır. Bir adet Namenode ve buna bağlı olan birden çok Datanode'dan oluşan HDFS'de master görevini Namenode yapmaktadır. Namenode tarafından gönderilen komutlar Datanode'lar tarafından alınır ve buna göre işlemler yapılır.

Dağıtık bir dosya sistemine sahip olan HDFS ile birden fazla sunucunun diski bir araya gelerek tek bir sanal disk oluşturulur. HDFS'de verilerin bir kopyası birden fazla düğümde kayıt altına alındığı için hata anında veri kaybı yaşama durumu olmamaktadır.

Hadoop birçok firma ve akademik grup tarafından büyük verileri analiz etmek amacı ile kullanılmaktadır. Güvenilir, ölçeklenebilir ve veriyi düşük bütçeler ile işleyebilir olması Hadoop'u popüler kılmaktadır. Veriler analiz edilirken tek makine yerine birden fazla makineye dağıtıldığından işin tamamlanma süresi minimuma inmektedir.

Hadoop, tek bir sunucu üzerinde çalışabildiği gibi, kendi CPU ve hafıza birimi bulunan binlerce sunucusu olan bir küme üzerinde de çalışabilir. Hadoop, Common, HDFS, YARN ve MapReduce olmak üzere dört modülden oluşur.

TABLO I HADOOP MODÜLLERİ

| Common | Tüm Hadoop modüllerini destekleyen ortak modül |
|---|---|
| HDFS | Dağıtık dosya sistemi |
| YARN | Kaynak yönetimi ve iş zamanlaması yapan kütüphane |
| MapReduce | Dağıtık, paralel hesaplama kütüphanesi |

Hadoop, GFS (Google File System) tabanlı geliştirilmiş olan bir Hadoop Dağıtık Dosya Sistemine (HDFS) sahiptir. HDFS dosyaların bir küme üzerinde dağıtılmalarından sorumludur [2].

Hadoop sisteminde veriler bloklara bölünür. Her bir bloğun büyüklüğü varsayılan olarak 64 MB olarak ayarlanmıştır. Farklı blok büyüklüğü dfs.block.size parametresiyle ayarlanabilir. Verilerin bloklara ayrılması 64 MB'lık veri büyüklüğü elde edilen satırdan sonra yeni bir blok oluşturarak gerçekleştirilir [3].

## C. MapReduce

MapReduce, büyük verileri işlemek için geliştirilmiş dağıtık bir programlama modelidir [4]. Dağıtık bir sistemde veriler MapReduce ile analiz edilirken iki fonksiyon kullanılır. Bunlardan birincisi map fonksiyonu diğeri de reduce fonksiyonudur [5].

MapReduce modelinin ilk adımı, elemanları bir anahtar ile eşleştirip bir sonraki işleme hazır hale getirir. İkinci adımda iteratör ile aynı anahtar için bir değerler listesi alır ve bunları biriktirip, filtreleyip, örnekleyip azaltır. Sonuçlarını da HDFS veya HBASE NoSQL veritabanına yazar[6].

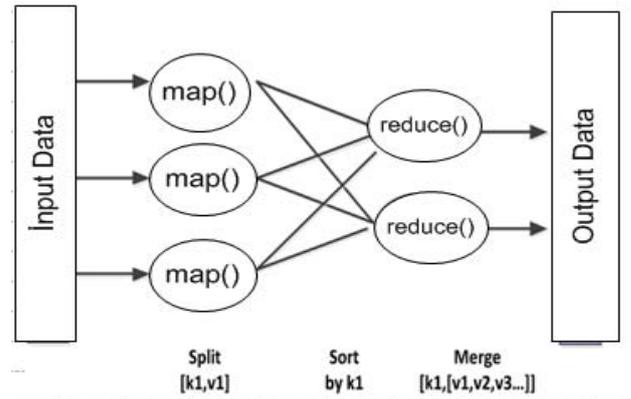

Şekil. 1. MapReduce [7]

## D. Apache Mahout

Apache Mahout, Hadoop platformu üzerinde, dağıtık bir şekilde ölçeklenebilir makine öğrenmesi algoritmalarını sunan Apache tarafından tasarlanmış olan bir projedir[8].

Öneri sistemleri, sınıflandırma ve kümeleme gibi işlemler, MapReduce kullanılarak ve dağıtık olarak Mahout ile kolaylıkla yapılabilmektedir. Mahout facebook, twitter, amazon gibi büyük firmalar tarafından günümüzde kullanılmaktadır. Mahout ile kümeleme yöntemlerini kullanarak verileri gruplayabilir ve ya sınıflandırma algoritmaları ile önceden isimlendirilmiş veriler üzerinden karar ağaçları oluşturulabilir.

## E. Bulut Bilişim

Kurulum gereksinimi olmadan her yerde çalışma desteği sunan bulut bilişim hizmeti, bilgiyi çevrimiçi olarak dağıtırken aynı zamanda gerekli yazılımları da paylaşarak mevcut olan hizmetlerin bir ağ üzerinden kullanılmasını sağlar. Elektronik posta, bulut bilişim için tipik bir örnek olup evlerde veya işyerlerinde bir yapılandırma olmaksızın bu hizmetten faydalanılmaktadır. Bunun dışında bulut depolama hizmetine örnek olarak Dropbox, GoogleDrive, iCloud verilebilir.

Kullandığın kadar öde mantığıyla çalışan çeşitli ticari bulutlar mevcut olup bunlara örnek olarak Google Cloud, Amazon EC2, GoGrid, FlexiScale verilebilir. Buna ek olarak, OpenStack, Eucalyptus, OpenNebula, ve Nimbus gibi açık kaynaklı bulut bilişim yazılımları kullanarak özel bulutlar oluşturmak mümkündür [9].

## F. Zemberek

Açık kaynak kodlu doğal dil işleme kütüphanesi Zemberek bir OpenOffice ve LibreOffice eklentisi olup Java ile geliştirilmiştir ve platform bağımsızdır [10].

Kelime gövdelerinin bulunması, gereksiz kelimelerin bulunup temizlenmesi için kullanılan Zemberek ile kelime kök ve gövdeleri, edatlar, bağlaçların tespiti gibi dil bilgisi gerektiren işlemler otomatik bir şekilde yapılır ve dokümanlar bu ön işlemlerden sonra hesaplama ortamına aktarılmaya hazır hale gelmiş olur.

### III. DAĞITIK DOKÜMAN SINIFLANDIRMA UYGULAMASI

Dijital dokümanların sayılarının gün geçtikçe çok hızlı bir şekilde artmasıyla birlikte dokümanların önceden belirlenmiş sınıflara ayrılması son yıllarda oldukça önem kazanmış bir konu olmuştur. Doküman sınıflandırma yöntemleri ile dokümanlar kategorilere, başlıklara ayrılabilir, spam (önemsiz) olup olmadığı, hangi dilde yazıldığı, hangi duygu durumunu gösterdiği tespit edilebilir.

Uygulama yapılacak olan platformlarda Hadoop kümeleri kurulumları yapıldıktan sonra ilk olarak Google Cloud üzerinde, Tübitak Dergipark'tan çekilmiş olan akademik makalelerin PDF dosyalarından metinlerin çıkarılması işlemi gerçekleştirilip, bu metinlerin hangi dilden olduğu tespit edilmiştir. Metinler öncelikle MongoDB NoSQL veritabanında kaydedilmiştir. PDF dosyalarından metin çıkarımı için Apache Tika kullanılmıştır

#### A. Apache Mahout ile Naive Bayes Sınıflandırma

PDF dokümanlarından çıkarılmış olan içerikler doğal dil işleme kütüphanesi olan Zemberekten geçirilerek gereksiz kelimeler (stop-word) atılmıştır. MongoDB'de tutulan metinler Hadoop üzerinde işlenebilmek için HDFS'e atılmıştır. TÜBİTAK Dergipark sitesinde verilen kategorilerin her birisi bir sınıf olarak kabul edilmiş ve böylece beş sınıf elde edilmiştir.

Sınıflandırılmamış dergiler veri setine eklenmemiştir. HDFS'e veriler atılırken her yayın kendi ismindeki klasörün altına kopyalanmıştır. HDFS'te var olan bir klasörün altına *mühendislik, hukuk, yaşam, sosyal ve tıp* olmak üzere beş adet sınıfı temsil eden klasörler ve onların altında da o sınıfa ait yayınlar kaydedilmiştir.

Bu şekilde depolanan dokümanlar üzerinde, aşağıda gösterilen komutlar kullanılarak Naive Bayes sınıflandırma yapılmıştır [11]:

```
./mahout seqdirectory -i /academic -o /academic-seq

./mahout seq2sparse -i /academic-seq -o /academic-vectors -lnorm -nv -wt tfidf

./mahout split -i /academic-vectors/tfidf-vectors --trainingOutput /academic-train-vectors --testOutput /academic-test-vectors --randomSelectionPct 40 --overwrite --sequenceFiles -xm sequential

./mahout trainnb -i /academic-train-vectors -el -o /academic-model -li /academic-labelindex -ow -c

./mahout testnb -i /academic-test-vectors -m /academic-model -l /academic-labelindex -ow -o /academic-testing –c
```

Toplam verinin %40'ı olan 19066 adet veri test verisi olarak kullanılmış, 28599 tanesi de eğitim verisi olarak ayrılmıştır. Google Cloud üzerinde çalıştırılan programda %89 doğruluk oranı gözlemlenmiştir.

```
=======================================================
Summary
-------------------------------------------------------
Correctly Classified Instances     :    17116    89.7724%
Incorrectly Classified Instances   :     1950    10.2276%
Total Classified Instances         :    19066

=======================================================
Confusion Matrix
-------------------------------------------------------
a      b      c      d      e      <--Classified as
2927   64     89     14     185    |  3279    a    = engineering
4      1      2      0      34     |  41      b    = law
81     213    5473   251    381    |  6399    c    = life
60     80     56     4879   364    |  5439    d    = medicine
17     17     9      29     3836   |  3908    e    = social

=======================================================
Statistics
-------------------------------------------------------
Kappa                                       0.8611
Accuracy                                   89.7724%
Reliability                                60.8491%
Reliability (standard deviation)            0.4638
Weighted precision                          0.9222
Weighted recall                             0.8977
Weighted F1 score                           0.9064

15/11/23 23:14:38 INFO MahoutDriver: Program took 52452 ms
(Minutes: 0.8742)
```

İkinci olarak, Amazon Cloud üzerinde 3 adet düğümden oluşan bir Hadoop kümesi üzerinde Apache Mahout ile Spark kullanılarak Naive Bayes Sınıflandırma yapılmıştır. Komutlar çalıştırılana kadarki verilerin içeriklerinin çıkarılması, zemberekten geçirilmesi, her yayının isminin olduğu klasöre atılması gibi yukarıda anlatılanlar ile aynı olmak üzere aşağıdaki komutlar çalıştırılarak sonuç elde edilmiştir:

```
mahout seq2sparse -i /file.seq -o /spark-sparse -
lnorm -nv -wt tfidf

mahout split -i /spark-sparse/tfidf-vectors --
trainingOutput /spark-training --testOutput /spark-
test --randomSelectionPct 40 --overwrite --
sequenceFiles -xm sequential

mahout spark-trainnb -i /spark-training -o /spark-
model -ow

mahout spark-testnb -i /spark-test -m /spark-model
```

Sonuçlar %91 doğru sınıflandırma oranına ulaşıldığını göstermektedir.

```
=================================================
Summary
-------------------------------------------------
--
Correctly Classified Instances: 17445 91.4836%
Incorrectly Classified Instances: 1624 8.5164%
Total Classified Instances: 19069

=================================================
Confusion Matrix
-------------------------------------------------
--
d    c    a    b    e        <--Classified as
4941 85   98   10   326  |   5460   d = medicine
214  5770 106  19   299  |   6408   c = life
17   102  2986 9    163  |   3277   a = engineeering
0    1    4    9    29   |   43     b = law
43   16   39   44   3739 |   3881   e = social

=================================================
Statistics
-------------------------------------------------
Kappa:                 0.8832
Accuracy:              91.4836%
Reliability:           64.8216%
Reliability (std dev): 0.4268
Weighted precision:    0.9218
Weighted recall:       0.9148
Weighted F1 score:     0.9166
```

TABLO II BEŞ KATEGORİ İLE SINIFLANDIRMA İÇİN TEST SÜRELERİ

| Kümedeki Makine Sayısı | Test Süreleri (ms) |
|---|---|
| 4 | 114577 |
| 5 | 109275 |
| 6 | 90721 |
| 7 | 99973 |

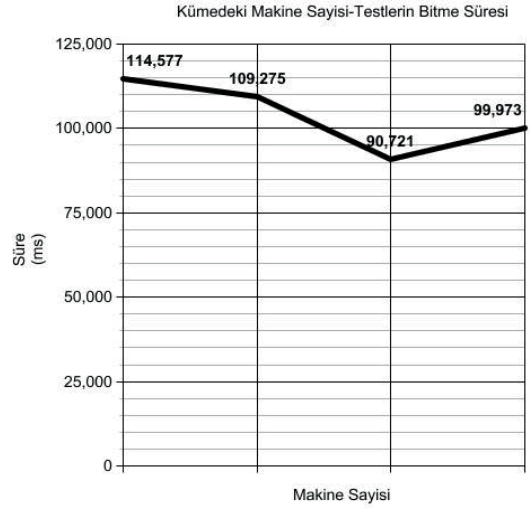

Şekil. 2. Makine sayısı- Bitme süresi

Sonuçlardan görüleceği gibi, 3 düğümden oluşan küme üzerinde Apache Mahout ile Naive Bayes Sınıflandırma yapılıp yüksek başarı oranı elde edilmiştir. Aynı şekilde Google Cloud üzerinde bu kez makine sayıları değiştirilerek testler yapılmıştır. 4 makine için testlerin bitme süresi 114577ms iken makine sayısı arttıkça testlerin bitme süresinin kısaldığı görülmüştür.

TABLO III : TEST VERİ MİKTARINA GÖRE SINIFLANDIRMA BAŞARI TABLOSU

| Test veri seti (%) | Başarı Oranı (%) |
|---|---|
| 10 | 88.0504 |
| 20 | 88.1862 |
| 30 | 89.574 |
| 40 | 89.7724 |

Apache Mahout kullanılarak, test veri setinin tüm veri setine oranı %40 iken sınıflandırma yapıldığında %89.7724 başarı oranı elde edilmiştir. Test veri seti yüzdesi değiştirildiğinde başarı oranının da değiştiği gözlenmiştir. Örneğin Tablo 4.6'da test veri seti %10 iken başarı oranı %88.050 olmuş, %20 iken başarı oranı %88.1862 ve %30 iken %89.574 şeklinde test veri seti oranı arttıkça başarı oranının da arttığı sonucuna ulaşılmıştır

## IV. SONUÇLAR

Bu çalışmada dağıtık makine öğrenmesi algoritmasının büyük verilere uygulanarak akademik makalelerin sınıflandırılması çalışması özetlenmiştir. Bunun için açık kaynak kodlu yazılımlar incelenip yeni ve popüler teknolojiler ile bunların veriler üzerine uygulanması yöntemleri üzerinde durulmuştur. Bulut üzerinde verileri işlemek için Apache Mahout teknolojisi ile uygulama gerçekleştirilmiştir.

Sağlık, sosyal, mühendislik, hukuk ve tıp olmak üzere beş kategoriye ayrılmış olan dokümanların Apache Mahout kullanılarak, Naive Bayes algoritması ile sınıflandırılması gerçekleştirilmiştir. Yapılan sınıflandırma ile kategorilerin tahmini ve performans anlamında yüksek başarı oranı elde edildiği gözlenmiştir.